\documentclass[conference]{IEEEtran}
\usepackage{algorithm}
\usepackage{graphicx}
\usepackage[noend]{algpseudocode}
\usepackage{etex} 
\usepackage{tikz}
\usetikzlibrary{shapes,arrows}
\usepackage{pstricks}
\usepackage{pst-node,pst-blur}
\usetikzlibrary{arrows}
\usepackage{amsfonts}
\usepackage{tabularx}
\usepackage{subfigure}
\usepackage{amssymb}
\usepackage{booktabs}
\usepackage{multirow}
\usepackage{epsfig}
\usepackage{epstopdf}
\usepackage[absolute]{textpos}
\usepackage{array}

\usepackage[noadjust]{cite}
\newtheorem{theorem}{Theorem}

\newtheorem{proposition}{Proposition}
\newtheorem{lemma}{Lemma}

\usepackage[cmex10]{amsmath}
\usepackage[cmex10]{mathtools}

\allowdisplaybreaks  
   
 \makeatletter 
 \def\@eqnnum{{\normalsize \normalcolor (\theequation)}} 
  \makeatother
\begin{document} 
\title{Harvested Power Maximization in QoS-Constrained MIMO SWIPT with Generic RF Harvesting Model} 
\author{\IEEEauthorblockN{Deepak Mishra$^{1}$ and George~C.~Alexandropoulos$^{2}$}
\IEEEauthorblockA{$^{1}$Department of Electrical Engineering, 
Indian Institute of Technology Delhi, India\\
$^{2}$Mathematical and Algorithmic Sciences Lab, Paris Research Center, Huawei Technologies France SASU, France\\
emails: dph.mishra@gmail.com, george.alexandropoulos@huawei.com
}} 
\bstctlcite{IEEEexample:BSTcontrol}

\begin{textblock}{14}(1,.1)
	\begin{center}
		Paper accepted to IEEE CAMSAP 2017.
	\end{center}
\end{textblock}

\maketitle  
\begin{abstract}
We consider the problem of maximizing the harvested power in Multiple Input Multiple Output (MIMO) Simultaneous Wireless Information and Power Transfer (SWIPT) systems with power splitting reception. Different from recently proposed designs, we target with our novel problem formulation at the jointly optimal transmit precoding and receive uniform power splitting (UPS) ratio maximizing the harvested power, while ensuring that the Quality-of-Service (QoS) requirement of the MIMO link is satisfied. We assume generic practical Radio Frequency (RF) Energy Harvesting (EH) receive operation that results in a non-convex optimization problem for the design parameters, which we then solve optimally after formulating it in an equivalent generalized convex form. Our representative results including comparisons of achievable EH gains with benchmark schemes provide key insights on various system parameters.  
\end{abstract}

%\begin{IEEEkeywords}
%RF energy harvesting, multiple input multiple output, optimization, precoding, power splitting, rate-energy trade off, simultaneous wireless information and power transfer.
%\end{IEEEkeywords}

\section{Introduction}\label{sec:introduction}  
There has been recently increasing interest~\cite{SWIPT_modern,Niyato_Survey1,ComMag} in utilizing Radio Frequency (RF) signals for Simultaneous Wireless Information and Power Transfer (SWIPT). Although SWIPT technology in conjunction with the adoption of wireless devices capable of performing Energy Harvesting (EH) is one of the promising candidates for enabling the perpetual operation of wireless devices, it suffers from some fundamental bottlenecks. Firstly, the signal processing and resource allocation strategies for efficient Information Decoding (ID) and EH differ significantly in their respective goals~\cite{SMT}. In addition, SWIPT performance is impacted by the low energy sensitivity and RF-to-Direct Current (DC) rectification efficiency~\cite{ComMag}. Another practical problem with SWIPT is the fact that the existing RF EH circuits cannot decode the information directly, and vice-versa~\cite{Tradeoff}. Lastly, the available solutions~\cite{MIMO_SWIPT,MISO-PS,MUTxMIMO,ASUPS} for realizing practically achievable SWIPT gains require high complexity and are still far from providing analytical insights. To confront with these bottlenecks, the Multiple-Input-Multiple-Output (MIMO) technology and joint resource allocation schemes have been recently considered~\cite{DPS,Tradeoff,MIMO_SWIPT,MISO-PS,TCOM2,MUTxMIMO,ASUPS,Our_GC2017}.

The non-trivial trade off between information capacity and average received power {for EH} was firstly investigated in~\cite{SMT}. In~\cite{Tradeoff}, the authors discussed why SWIPT theoretical gains are difficult to realize, and proposed some practical Receiver (RX) {architectures}. These include Time Switching (TS), Power Splitting (PS), and Antenna Switching architectures~\cite{DPS} that use one portion of the received signal (in time, power, or space) for EH and the other one for ID. The aforementioned RX architectures for SWIPT have been lately considered in various MIMO system setups~\cite{MIMO_SWIPT,MISO-PS,MUTxMIMO,ASUPS}. Transmitter (TX) precoding techniques for efficient SWIPT in RF-powered MIMO systems were presented in~\cite{MIMO_SWIPT}. In \cite{MISO-PS}, a Semi-Definite Programming (SDP) relaxation technique for a multi-user multiple-input single-output system was used to study the joint TX precoding and PS optimization for minimizing the transmit power under signal-to-interference-plus-noise ratio and EH constraints. In~\cite{MUTxMIMO} and~\cite{ASUPS}, more general MIMO interference channels were investigated adopting the interference alignment technique. However, the vast majority of the available MIMO SWIPT studies presented suboptimal iterative algorithms for the design parameters that are based on either convex relaxation or approximation approaches. 

A major goal of RF EH systems is the optimization of the end-to-end EH efficiency~\cite{Niyato_Survey1}. This is in principle challenging because the RF-to-DC rectification is a non-linear function of the received RF power~\cite{NonRFH,ICC17Wksp,TCOM}. This corroborates the necessity of optimizing the harvested power rather than the receiver power, which is usually treated in the existing literature~\cite{MIMO_SWIPT,DPS,MISO-PS,MUTxMIMO,ASUPS,TCOM2}; therein, constant RF-to-DC rectification efficiency was assumed. In this paper, we aim at maximizing the harvested power in MIMO SWIPT systems with PS reception, while ensuring that the underlying Quality-of-Service (QoS) expressed in terms of a minimum rate requirement of the MIMO link is satisfied. We adopt a generic practical RF EH model and present a novel global jointly optimal TX precoding and RX Uniform PS (UPS) design. Our selected numerical investigations provide key insights on the interplay among various system parameters on the optimized trade off between the harvested power and the underlying QoS requirement. 

\textit{Notations:} Vectors and matrices are denoted by boldface lowercase and boldface capital letters, respectively. $\mathbf{A}^{\rm H}$ and  $\mathrm{tr}\left(\mathbf{A}\right)$ respectively denote Hermitian transpose and trace of matrix $\mathbf{A}$, while $\mathbf{I}_{n}$ is the $n\times n$ identity matrix.  $[\mathbf{A}]_{i,j}$ stands for $\mathbf{A}$'s $(i,j)$-th element, $\lambda_{\max}\left(\mathbf{A}\right)$ represents the largest eigenvalue of $\mathbf{A}$, and ${\rm diag}\{\cdot\}$ denotes a square diagonal matrix with $\mathbf{a}$'s elements in its main diagonal. $\mathbf{A}^{-1}$ represents the inverse of a square matrix $\mathbf{A}$, whereas $\mathbf{A}\succeq0$ means that $\mathbf{A}$ is positive semi definite. $\mathbb{C}$ represents the complex number set, $\left(x\right)^+\triangleq\max\left\{0,x\right\}$, and $\mathbb{E}\{\cdot\}$ is the expectation operator.
  
\section{System and Channel Models}\label{sec:system_model}  
We consider a MIMO SWIPT system  where the TX is equipped with $N_T$ antenna elements and wishes to simultane-

\noindent ously transmit information and energy to the RF EH RX having $N_R$ antenna elements. We assume a frequency flat MIMO fading channel $\mathbf{H}\in\mathbb{C}^{N_R\times N_T}$ that is perfectly known at both TX and RX. The entries of $\mathbf{H}$ are assumed to include independent, Zero-Mean Circularly Symmetric Complex Gaussian (ZMCSCG) random variables with unit variance;  hence the rank of $\mathbf{H}$ is $r=\min(N_R,N_T)$. The baseband received signal $\mathbf{y}\in\mathbb{C}^{N_R\times 1}$ at RX is given by $\mathbf{y}\triangleq\mathbf{H}\mathbf{x} + \mathbf{n}$, where $\mathbf{x}\in\mathbb{C}^{N_T\times 1}$ denotes the transmitted signal with covariance matrix $\mathbf{S}\triangleq\mathbb{E}\{\mathbf{x}\mathbf{x}^{\rm H}\}$ and $\mathbf{n}\in\mathbb{C}^{N_R\times 1}$ represents the Additive White Gaussian Noise (AWGN) vector having ZMCSCG entries each with variance $\sigma^2$. The elements of $\mathbf{x}$ and $\mathbf{n}$ are assumed to be statistically independent. For the transmitted signal we finally assume that there exists an average power constraint $\mathrm{tr}\left(\mathbf{S}\right)\le P_T$ across all TX antennas.

Capitalizing on the latter signal model, the {average} received power $P_R$ across all RX antennas can be obtained as $P_R=\mathbb{E}\{\mathbf{y}^{\rm H}\mathbf{y}\}$; the latter averaging is performed over the transmitted symbols during each coherent channel block. {As} the noise strength (generally lower than $-80$dBm) is much below than the received energy sensitivity of practical RF {EH} circuits (which is around $-20$dBm)~\cite{Niyato_Survey1}, we next neglect the contribution of $\mathbf{n}$ to the harvested power. Note, however, that the analysis and optimization results of this paper can be easily extended for non-negligible noise power scenarios. We therefore rewrite $P_R$ as the following function of $\mathbf{H}$ and $\mathbf{x}$:
\begin{equation}\label{eq:PR}
P_R\triangleq\mathbb{E}\left\{\mathbf{x}^{\rm H}\mathbf{H}^{\rm H}\mathbf{H}\mathbf{x}\right\}=\mathrm{tr}\left(\mathbf{H}\mathbf{S}\mathbf{H}^{\rm H}\right).
\end{equation} 
We consider the UPS ratio $\rho\in[0,1]$ at each RX antenna element, which reveals that $\rho$ fraction of the received signal power at each antenna is used for RF EH, while the remaining $1-\rho$ fraction is used for ID. With this setting, the average total received power $P_{R,E}$ available for RF EH is given by $P_{R,E}\triangleq\rho P_R=\rho\,\mathrm{tr}\left(\mathbf{H}\mathbf{S}\mathbf{H}^{\rm H}\right)$~\cite{MIMO_SWIPT}. Supposing that $\eta\left(\cdot\right)$ denotes the RF-to-DC rectification efficiency function, which is in general a non-linear positive function of the received RF power $P_{R,E}$ available for EH~\cite{ICC17Wksp,NonRFH}, the total harvested DC power is obtained as $P_H\triangleq\eta\left(\rho P_R\right)\rho P_R$. Despite this non-linear relationship between $\eta$ and $P_{R,E}$, $P_H$ is monotonically non-decreasing in $P_{R,E}=\rho P_R$ for any practical RF EH circuit~\cite{NonRFH}, \cite[Fig. 3]{ICC17Wksp} due to the law of energy conservation.  
 
\section{Joint Optimization Formulation}\label{sec:problem}
Focusing on the MIMO SWIPT model of Section~\ref{sec:system_model}, we consider the problem of designing the covariance matrix $\mathbf{S}$ at the multi-antenna TX and the UPS ratio $\rho$ at the multi-antenna RF EH RX for maximizing the total harvested DC power, while satisfying a minimum instantaneous rate requirement ${R}$ in bits per second per Hz (bps/Hz) for information transmission. In mathematical terms, the considered joint design optimization problem is expressed as:
\begin{equation*}\label{eqOPT}
\begin{split}
  \mathcal{OP}: &\max_{\rho,\mathbf{S}} \quad P_H=\eta\left(\rho\,\mathrm{tr}\left(\mathbf{H}\mathbf{S}\mathbf{H}^{\rm H}\right)\right)\rho\,\mathrm{tr}\left(\mathbf{H}\mathbf{S}\mathbf{H}^{\rm H}\right) \quad \textrm{s.t.}
	\\&({\rm C1}):~\log_2\left(\det\left(\mathbf{I}_{N_R}+\left(1-\rho\right)\sigma^{-2}\mathbf{H}\mathbf{S}\mathbf{H}^{\rm H}\right)\right)\ge{R}, 
	\\&({\rm C2}):~\mathrm{tr}\left(\mathbf{S}\right)\leq P_T,~({\rm C3}):~\mathbf{S}\succeq 0,~({\rm C4}):~0\leq\rho\leq1,
\end{split}
\end{equation*}
where constraint $({\rm C1})$ represents the minimum instantaneous

\noindent rate requirement, $({\rm C2})$ is the average transmit power constraint, while constraints $({\rm C3})$ and $({\rm C4})$ are the boundary conditions for $\mathbf{S}$ and $\rho$. It can be easily concluded from $\mathcal{OP}$ that the objective function $P_H$ is jointly non-concave in regards to the unknown variables $\mathbf{S}$ and $\rho$. It will be shown, however, in the following Lemma~\ref{lem:pcc} that the received power $P_{R,E}$ available for EH is jointly pseudoconcave in $\mathbf{S}$ and $\rho$.  
\begin{lemma}\label{lem:pcc}
The received RF power $P_{R,E}$ available for EH is a joint pseudoconcave function of $\mathbf{S}$ and $\rho$.
\end{lemma} 
\begin{IEEEproof}
As $\mathrm{tr}\left(\mathbf{H}\mathbf{S}\mathbf{H}^{\rm H}\right)$ is linear in $\mathbf{S}$, we deduce that $P_{R,E}=\rho\,\mathrm{tr}\left(\mathbf{H}\mathbf{S}\mathbf{H}^{\rm H}\right)$ available for EH is the product of two positive concave functions of $\rho$ and $\mathbf{S}$. Since the product of two positive concave functions is log-concave~\cite[Chapter 3.5.2]{boyd} and a positive log-concave function is also pseudoconcave~\cite[Lemma 5]{TCOM2}, $P_{R,E}$ is jointly pseudoconcave in $\mathbf{S}$ and $\rho$.
\end{IEEEproof}

We now show that solving $\mathcal{OP}$ is equivalent to solving the following joint optimization problem:
\begin{equation*}\label{eqOPT1}
  \mathcal{OP}1: \max_{\rho,\mathbf{S}} \quad P_{R,E}=\rho\,\mathrm{tr}\left(\mathbf{H}\mathbf{S}\mathbf{H}^{\rm H}\right)~~
	\textrm{s.t.}~~({\rm C1})-({\rm C4}). 
\end{equation*}
\begin{proposition}\label{prep}
The solution of $\mathcal{OP}1$ solves $\mathcal{OP}$ optimally.
\end{proposition} 
\begin{IEEEproof}
Recall that for any practical RF EH circuit, $P_H$ is always monotonically non-decreasing in $P_{R,E}$~\cite{NonRFH,ICC17Wksp}. It can be thus concluded \cite{avriel2010generalized,boyd} that the monotonic non-decreasing transformation $P_H$ of the pseudoconcave  $P_{R,E}$ is pseudoconcave and possesses the unique global optimality property~\cite[Props$.$ 3.8 and 3.27]{avriel2010generalized}. $\mathcal{OP}$ and $\mathcal{OP}1$ are thus equivalent~\cite{Baz} sharing the same solution pair $\left(\mathbf{S}^*,\rho^*\right)$.    
\end{IEEEproof}
Although $\mathcal{OP}1$ is non-convex, we next prove a property that will {be used} in Section~\ref{sec:soln} to derive its optimal solution. 
\begin{theorem}\label{th:Gen_Cvx}
$\mathcal{OP}1$ is a generalized convex problem and its globally optimal solution $\left(\mathbf{S}^*,\rho^*\right)$ can be obtained by solving its Karush-Kuhn-Tucker (KKT) conditions.
\end{theorem} 
\begin{IEEEproof}
From Lemma~\ref{lem:pcc}, $P_{R,E}$ is jointly pseudoconcave in $\mathbf{S}$ and $\rho$. In  $({\rm C1})$, $R-\log_2\left(\det\left(\mathbf{I}_{N_R}+\left(1-\rho\right)\sigma^{-2}\mathbf{H}\mathbf{S}\mathbf{H}^{\rm H}\right)\right)$ is a jointly convex function of $\rho$ and $\mathbf{S}$; this ensues from the fact that the matrix inside the determinant is a positive definite matrix. In addition, $({\rm C2})$ and $({\rm C3})$ are linear with respect to $\mathbf{S}$ and independent of $\rho$, whereas $({\rm C4})$ depends only on $\rho$ and is convex. The proof completes by combining the latter findings and using them in~\cite[Th. 4.3.8]{Baz}.        
\end{IEEEproof}

\section{Optimal TX Precoding and RX Power Splitting}\label{sec:soln}
Using the findings of Proposition~\ref{prep} and Theorem~\ref{th:Gen_Cvx}, we note that the joint TX precoding and UPS design for $\mathcal{OP}1$ will also result in the maximization of the harvested power $P_H^*=\eta\left(P_{R,E}^*\right)P_{R,E}^*$ in $\mathcal{OP}$ for any practical RF EH circuitry. To derive the global joint optimal design, we next investigate the fundamental trade off between TX Energy Beamforming (EB) and information Spatial Multiplexing (SM) {in $\mathcal{OP}1$}. 

\subsection{Energy Beamforming versus Spatial Multiplexing Trade off}\label{sec:tradeoff}
Let us first consider the reduced Singular Value Decomposition (SVD) of $\mathbf{H}=\mathbf{U}\boldsymbol{\Lambda}\mathbf{V}^{\rm H}$, where $\mathbf{V}\in\mathbb{C}^{N_T\times r}$ and $\mathbf{U}\in\mathbb{C}^{N_R\times r}$ are unitary matrices and $\boldsymbol{\Lambda}\in\mathbb{C}^{r\times r}$ is the diagonal matrix with the of $r$ non-zero eigenvalues of $\mathbf{H}$ in decreasing order of magnitude. Ignoring $({\rm C1})$ in $\mathcal{OP}1$ leads to the rank-$1$ optimal TX covariance matrix $\mathbf{S^*}=\mathbf{S_{_\mathrm{EB}}}\triangleq P_T\,\mathbf{v}_1\mathbf{v}_1^{\rm H}$~\cite{MIMO_SWIPT}, where $\mathbf{v}_1\in\mathbb{C}^{N_T\times 1}$ is the first column of $\mathbf{V}$ that corresponds to the eigenvalue $[\boldsymbol{\Lambda}]_{1,1}\triangleq\lambda_{\max}\left(\mathbf{H}^{\rm H}\mathbf{H}\right)$. This TX precoding called TX EB maximizes the average received power by allocating $P_T$ to the strongest eigenmode of $\mathbf{H}^{\rm H}\mathbf{H}$.  On the other hand, it is well known \cite{MIMO_WF} that SM using waterfilling maximizes the information rate $R$ by performing optimal allocation of $P_T$ over the available eigenchannels of $\mathbf{H}$. Obviously, there exists a trade off between EB and SM that needs to be investigated in the framework of $\mathcal{OP}1$. 

\begin{figure}[!t]
\centering
{{\includegraphics[width=2.75in]{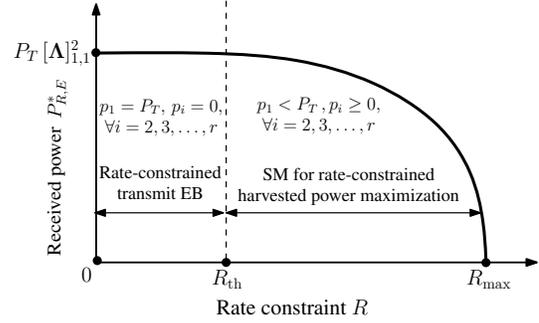} }} 
\caption{\small Illustration of the EB versus SM tradeoff and the threshold $R_{\mathrm{th}}$ that determines the optimal TX precoding operation.} 
    \label{fig:concept}
\end{figure}

Suppose we adopt EB in $\mathcal{OP}1$, then it results in the received RF power $P_{R_{_{\mathrm{EB}}}}\triangleq\rho_{_\mathrm{EB}}P_T\,[\boldsymbol{\Lambda}]_{1,1}^2$, where the UPS ratio $\rho_{_\mathrm{EB}}\triangleq\max\left\lbrace0,1-{\left(2^{{R}}-1\right) \sigma ^2}\left[{P_T\,[\boldsymbol{\Lambda}]_{1,1}^2}\right]^{-1}\right\rbrace$ is obtained by solving $({\rm C1})$ at equality. As $\rho_{_\mathrm{EB}}$  is a decreasing function of ${R}$,  there exists a rate threshold $R_{\rm th}$ such that, when $R>R_{\rm th}$, one should allocate $P_T$ over at least two eigenchannels instead of assigning $P_T$ solely to the strongest one. Consider the optimal power allocation $p_1^*$ and $p_2^*$ {for} the two highest gained eigenchannels with eigenmodes $[\boldsymbol{\Lambda}]_{1,1}$ and $[\boldsymbol{\Lambda}]_{2,2}$, respectively, with $[\boldsymbol{\Lambda}]_{1,1}>[\boldsymbol{\Lambda}]_{2,2}$. By substituting these values into $({\rm C1})$ and solving it at equality for $\rho$ gives the optimum UPS parameter $\rho_{_{\mathrm{SM}_2}}$ for SM over two eigenchannels. The resulting maximum received power for EH is given by $P_{R_{_{\mathrm{SM}_2}}}\triangleq\rho_{_{\mathrm{SM}_2}}\left(p_1^*\,[\boldsymbol{\Lambda}]_{1,1}^2+p_2^*\,[\boldsymbol{\Lambda}]_{2,2}^2\right)$. Combining the above along with $D_{12}\triangleq[\boldsymbol{\Lambda}]_{1,1}^2-[\boldsymbol{\Lambda}]_{2,2}^2$ yields that the $R_{\rm th}$ value rendering EB more beneficial than SM in terms of the RF power available for EH after meeting $({\rm C1})$ is given by
\begin{eqnarray}\label{eq:rth}
\textstyle R_{\mathrm{th}} =  \log_2\left(1+\frac{p_2^*D_{12}}{\sigma ^2}+\sqrt{\frac{D_{12} ([\boldsymbol{\Lambda}]_{1,1}^2 p_1^*+p_2^*[\boldsymbol{\Lambda}]_{2,2}^2)^2}{[\boldsymbol{\Lambda}]_{1,1}^2 [\boldsymbol{\Lambda}]_{2,2}^2 \sigma ^2 p_1^*}}\right).
\end{eqnarray} 
$R_{\mathrm{th}}$ evinces a \textit{switching point} on the desired TX precoding operation that is graphically presented in Fig$.$~\ref{fig:concept}. When the rate requirement $R\le R_{\mathrm{th}}$, TX EB is sufficient to meet $R$, and hence, can be used for maximizing the received RF power. For cases where ${R}>{R_{\mathrm{th}}}$, SM needs to be adopted for maximizing the received power for EH, while satisfying $R$. 

We next use this $R_{\mathrm{th}}$ definition (cf. \eqref{eq:rth}) to obtain the global jointly optimal TX precoding and RX UPS design for $\mathcal{OP}1$.

\subsection{Globally Optimal Solution of $\mathcal{OP}1$}\label{sec:KKT}
Associating the Lagrange multipliers $\mu$ and $\nu$ with $({\rm C1})$ and $({\rm C2})$, respectively, while {keeping} $({\rm C3}),({\rm C4})$ implicit, the Lagrangian function of $\mathcal{OP}1$ can be formulated as
\begin{align}\label{eq:Lang}
\mathcal{L}\triangleq&\rho\,\mathrm{tr}\left(\mathbf{HSH}^{\rm H}\right)-\nu\left(\mathrm{tr}\left(\mathbf{S}\right)-P_T\right)\nonumber
\\&\!\!-\mu\left({R}-\log_2\left[\det\left(\mathbf{I}_{N_R}+\left(1-\rho\right)\sigma^{-2}\mathbf{H}\mathbf{S}\mathbf{H}^{\rm H}\right)\right]\right).
\end{align}
By using Theorem~\ref{th:Gen_Cvx}, the globally optimal solution $\left(\mathbf{S^*},\rho^*\right)$ for $\mathcal{OP}1$ is obtained from the KKT conditions: $\frac{\partial \mathcal{L}}{\partial \mathbf{S}}=0$, $\frac{\partial \mathcal{L}}{\partial \rho}=0$, $ \mu\left({R}-\log_2\left[\det\left(\mathbf{I}_{N_R}+\frac{1-\rho}{\sigma^2}\mathbf{H}\mathbf{S}\mathbf{H}^{\rm H}\right)\right]\right)=0$, $\nu\left(\mathrm{tr}\left(\mathbf{S}\right)-P_T\right)=0$, and  $({\rm C1})$--$({\rm C4})$ along with $\mu,\nu\ge 0$. Since the available transmit power $P_T$ is always fully utilized due to the monotonically increasing nature of the objective $P_{R,E}$ in $\mathbf{S}$, $\nu\neq0$ and $\mathrm{tr}\left(\mathbf{S}\right)= P_T$ holds. Similarly, $\mu\neq0$ since $P_{R,E}$ is strictly increasing in $\rho$ and $1-\rho$ fraction of $P_{R}$ needs to be sufficient in meeting rate $R$ in {$({\rm C1})$}. 
 
Recalling the trade off discussion in Section~\ref{sec:tradeoff}, the optimal TX covariance matrix is $\mathbf{S_{_\mathrm{EB}}}$ with optimal UPS ratio $\rho_{_\mathrm{EB}}$ when $R\le{R_{\mathrm{th}}}$. By substituting $\mathbf{S_{_\mathrm{EB}}}$ and $\rho_{_\mathrm{EB}}$ into $\frac{\partial \mathcal{L}}{\partial \mathbf{S}}=0$ and $\frac{\partial \mathcal{L}}{\partial \rho}=0$, the optimal $\mu$ and $\nu$ are $\mu_{_\mathrm{EB}}\triangleq{\sigma^2\,\ln2}\left(1+ {\left(1-\rho_{_\mathrm{EB}}\right)P_T[\boldsymbol{\Lambda}]_{1,1}^2}{\sigma^{-2}}\right)$ and $\nu_{_\mathrm{EB}}\triangleq\frac{\mu_{_\mathrm{EB}}\left(1-\rho_{_\mathrm{EB}}\right)[\boldsymbol{\Lambda}]_{1,1}^2}{\ln(2)\left(\sigma^2+\left(1-\rho_{_\mathrm{EB}}\right)P_T[\boldsymbol{\Lambda}]_{1,1}^2\right)}+[\boldsymbol{\Lambda}]_{1,1}^2 \rho_{_\mathrm{EB}}$. Thus, the KKT point $\left(\mathbf{S^*},\rho^*,\mu^*,\nu^*\right)$ is given by $\left(\mathbf{S}_{_\mathrm{EB}},\rho_{_\mathrm{EB}},\mu_{_\mathrm{EB}},\nu_{_\mathrm{EB}}\right)$ for $R\le{R_{\mathrm{th}}}$. When $R>{R_{\mathrm{th}}}$, the optimal TX covariance matrix $\mathbf{S_{_\mathrm{SM}}}$ as obtained by solving $\frac{\partial \mathcal{L}}{\partial \mathbf{S}}=0$ in $ \mathbf{S}$ is given by
\begin{equation}\label{eq:optS0}
\textstyle\mathbf{S_{_\mathrm{SM}}}\triangleq\mathbf{V}\left(\frac{\mu_{_{\rm SM}}\left(\nu_{_{\rm SM}}\,\mathbf{I}_r-\rho_{_{\rm SM}}\,\boldsymbol{\Lambda}^{\rm H}\boldsymbol{\Lambda}\right)^{-1}}{\ln2}-\frac{\sigma^2\mathbf{\Lambda}^{-2}}{1-\rho_{_{\rm SM}}}\right)\mathbf{V}^{\rm H}.
\end{equation}
We rewrite \eqref{eq:optS0} as $\mathbf{S_{_\mathrm{SM}}}= \mathbf{V}\mathbf{P_{_\mathrm{SM}}}\mathbf{V}^{\rm H}$ in \eqref{eq:optS0}, where $\mathbf{P_{_\mathrm{SM}}}$ is the $r\times r$ diagonal power allocation matrix with its non-zero elements representing the optimal power allocation over the $r$ eigenchannels. The remaining three unknown parameters $\rho_{_{\rm SM}}, \mu_{_{\rm SM}}$, and $\nu_{_{\rm SM}}$ for SM when $R>{R_{\mathrm{th}}}$ are obtained by solving the system of three equations: $\log_2\left[\det\left(\mathbf{I}_{N_R}+\frac{1-\rho}{\sigma^2}\right.\right.$ $\left.\left.\!\!\mathbf{H}\mathbf{S}_{_\mathrm{SM}}\mathbf{H}^{\rm H}\right)\right]=R$, $\mathrm{tr}\left(\mathbf{S_{_\mathrm{SM}}}\right)=P_T$, and $\frac{\partial \mathcal{L}}{\partial \rho}=0$  after setting $\mathbf{S}=\mathbf{S_{_\mathrm{SM}}}$ and satisfying $\mu,\nu>0$ and $0\le\rho<1$. Hence, the KKT point for $R>{R_{\mathrm{th}}}$ is given by $\left(\mathbf{S}_{_{\rm SM}},\rho_{_{\rm SM}},\mu_{_{\rm SM}},\nu_{_{\rm SM}}\right)$.

We have thus obtained the jointlly optimal design $\left(\mathbf{S^*},\rho^*\right)$ for $\mathcal{OP}1$ that depends on the relative values of $R$ and ${R_{\mathrm{th}}}$. The feasibility of $\mathcal{OP}1$ depends on ${R_{\max}}\triangleq\log_2\left(\det\left(\mathbf{I}_{N_R}+\sigma^{-2}\mathbf{H}\mathbf{S_{_\mathrm{WF}}}\mathbf{H}^{\rm H}\right)\right)$ that represents the maximum achievable rate when the entire received signal strength is dedicated for ID (i$.$e$.$, $\rho=0$ and no EH). In this case, the TX covariance matrix is $\mathbf{S_{_\mathrm{WF}}}\triangleq \mathbf{V}\mathbf{P_{_\mathrm{WF}}}\mathbf{V}^{\rm H}$ with $\mathbf{P_{_\mathrm{WF}}}\triangleq{\rm diag}\{[p_{_{\mathrm{WF},1}}\, p_{_{\mathrm{WF},2}}\, \cdots\, p_{_{\mathrm{WF},r}}]\}$ being the power allocation matrix whose rank $r_w$ (non-zero diagonal entries) is given by
\begin{eqnarray}\label{eq:rnkWF}
\textstyle r_w\triangleq \max\left\lbrace k\middle|\left(P_T-\sum\limits_{i=1}^{k-1}\left(\frac{\sigma^2}{[\boldsymbol{\Lambda}]_{k,k}^2}-\frac{\sigma^2}{[\boldsymbol{\Lambda}]_{i,i}^2}\right)\right)^{\hspace{-1.5mm}+}\hspace{-2mm}>0, k\le r\hspace{-1mm}\right\rbrace.\hspace{-3mm}
\end{eqnarray} 
The $P_{_\mathrm{WF}}$'s non-zero diagonal elements are obtained using the standard waterfilling algorithm~\cite{WFG}.

\begin{figure*}[!t]
\centering
\subfigure[Trade off between $P_{R,E}^*$ and $R$ in $4\hspace{-0.2mm}\times\hspace{-0.2mm}4$ systems.]{{\includegraphics[width=2.35in]{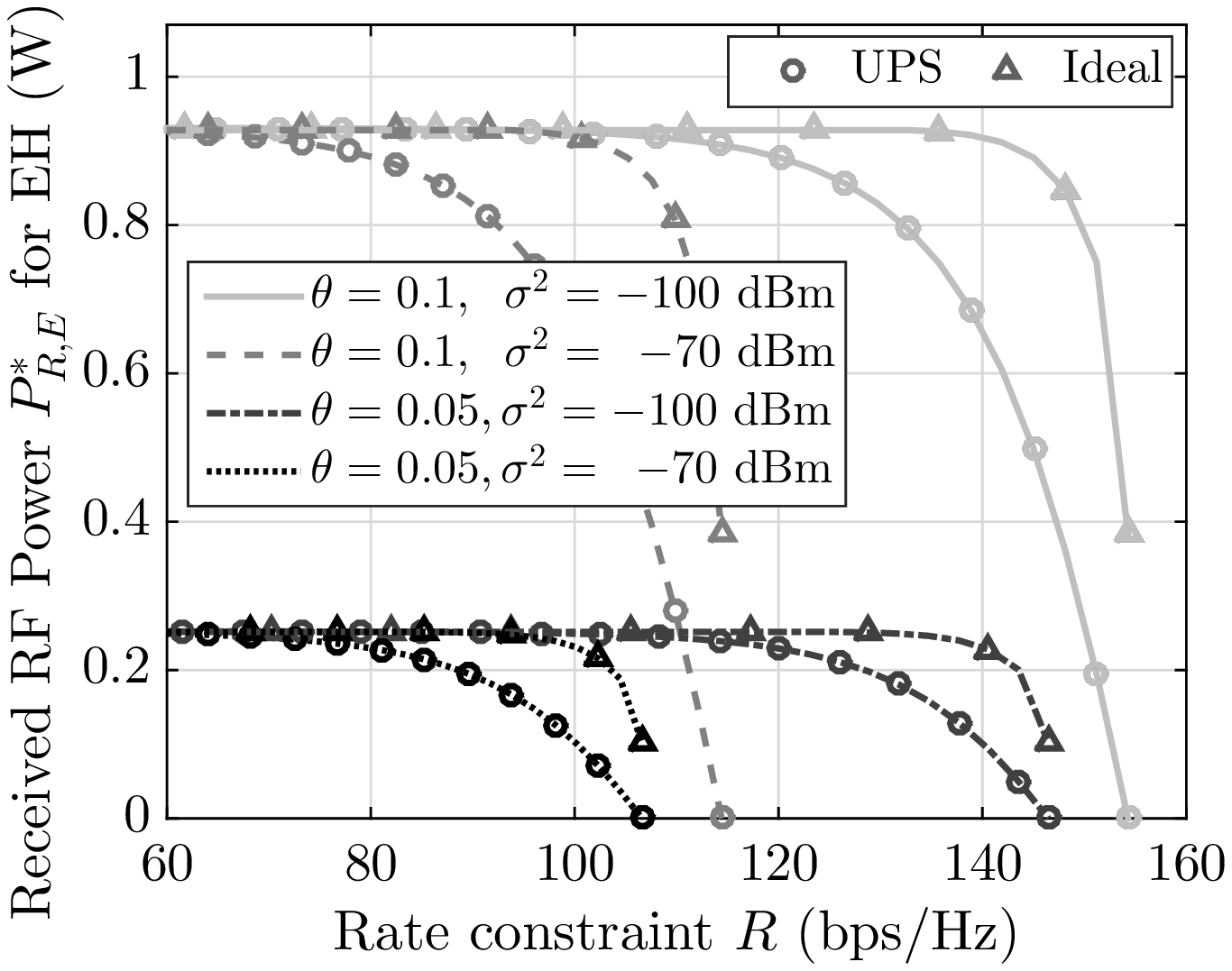} }}\hspace{-2mm}
\subfigure[Optimal power allocation in $4\times 4$ systems.]{{\includegraphics[width=2.35in]{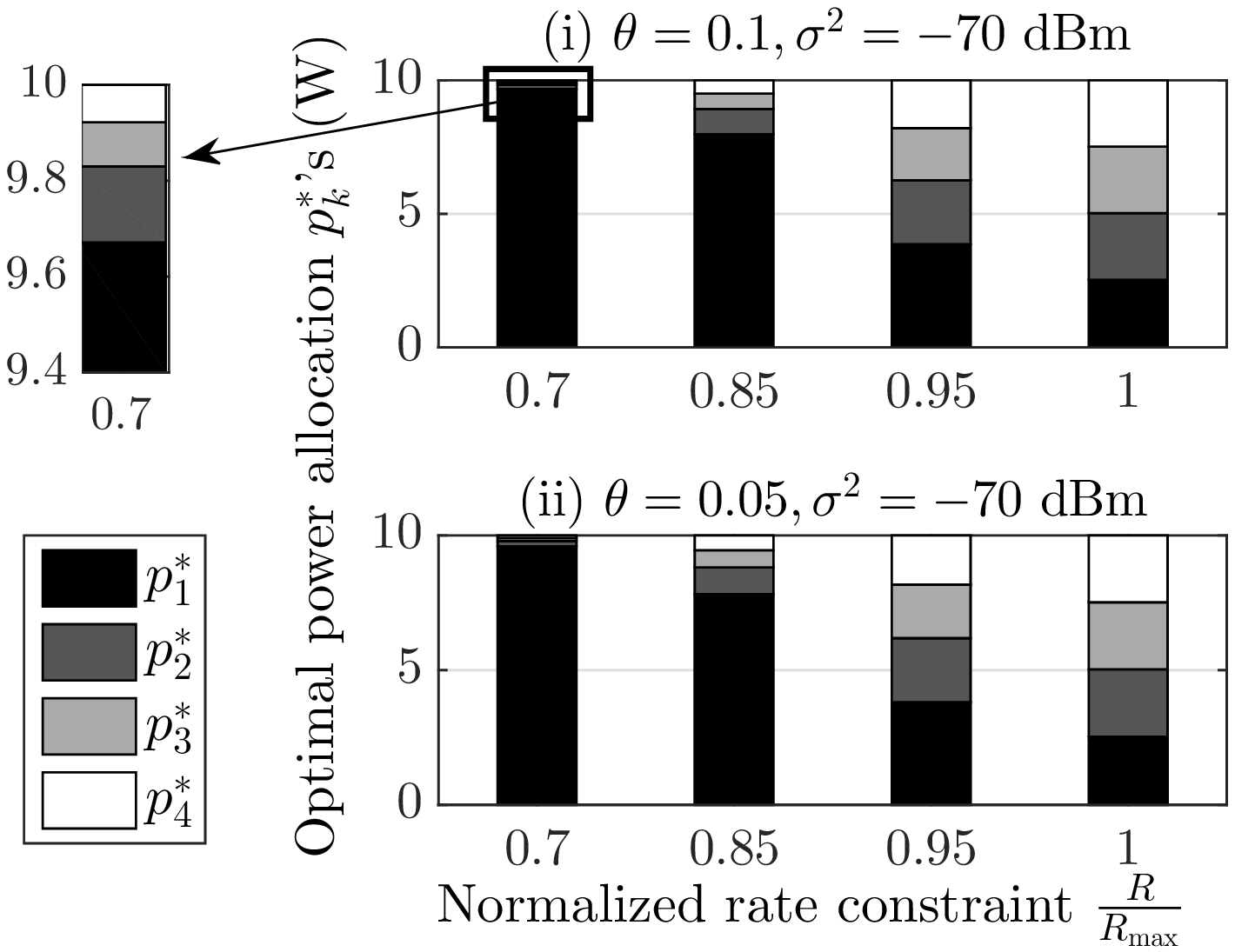} }}\,
\subfigure[Optimal UPS ratio in $2\times 2$ and $4\times 4$ systems.]{{\includegraphics[width=2.25in]{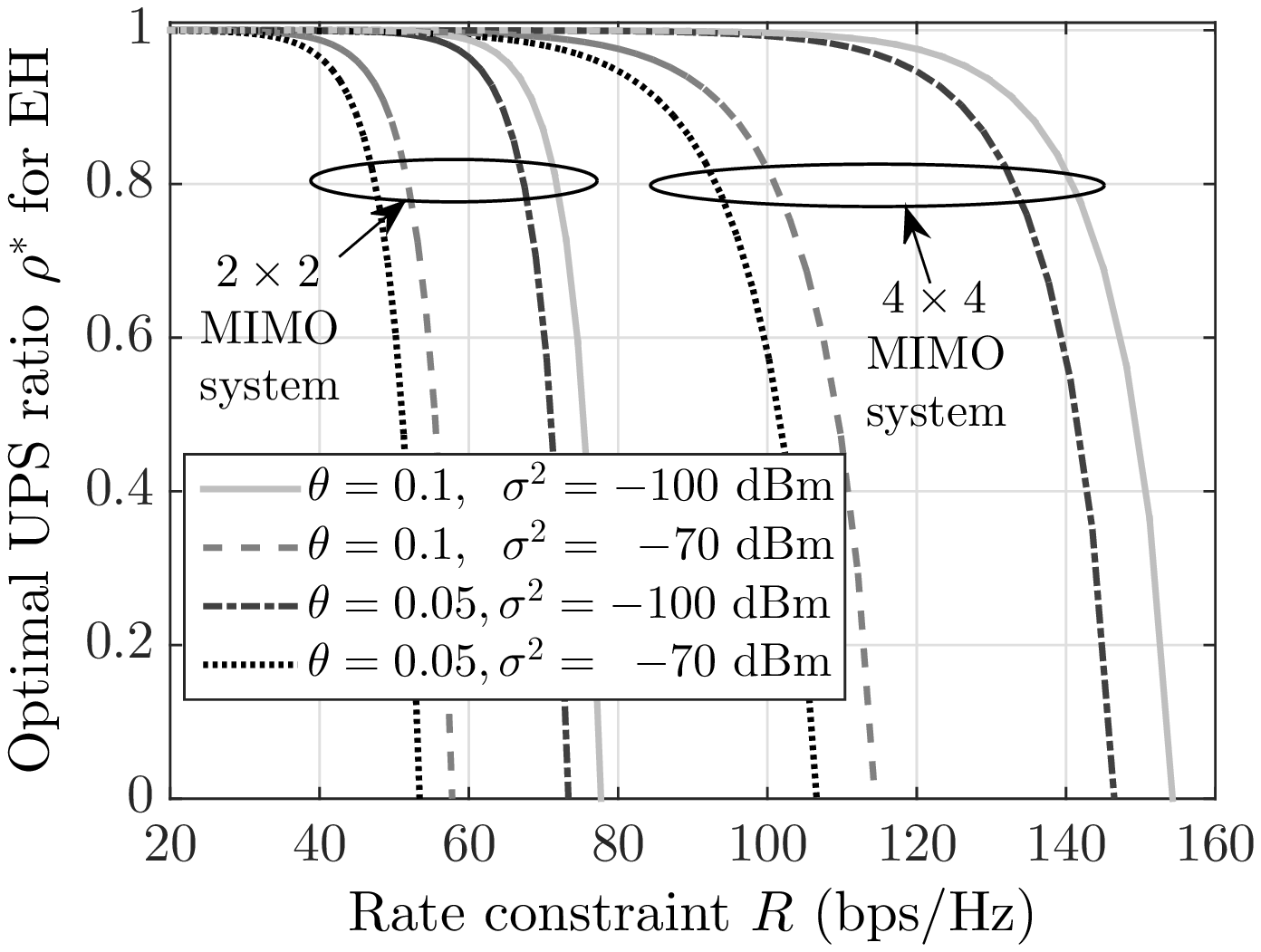} }} 
\caption{\small Variation of the optimized rate-energy tradeoff ($R$ versus $P_{R,E}^*$) and optimal design parameters for different values of $N$, $\theta$, and $\sigma^2$.} \label{fig:combined}
\end{figure*}

\section{Results and Discussion}\label{sec:results} 
We investigate the impact of various system parameters on the achievable rate-energy trade off with the proposed jointly optimal design. Unless otherwise stated, we set $P_T\!=\!10$W  and $\sigma^2\!=\!\{-100,-70\}$dBm by considering noise spectral density of $-175$ dBm/Hz. We set $\mathbf{H}=\big\{\theta h_{ij}\,\big| 1\le i,j\le N\big\}$ with $N\!\triangleq\! N_R\!=\!N_T$, where $\theta\!=\!\left\{0.1,0.05\right\}$ models propagation losses and $h_{ij}$'s are unit variance ZMCSCG random variables. Considering the unit transmission block duration assumption, we use the terms `received energy' and `received power' interchangeably. All performance results have been generated after averaging over $10^3$ independent channel realizations.
  
In Fig$.$~\ref{fig:combined}(a) we illustrate the optimized rate-energy trade off in a $4\times 4$ MIMO system for different values of $\theta$ and $\sigma^2$. As expected, lesser noisy systems when $\sigma^2$ decreases, and better channel conditions with increasing $\theta$ result in better trade off and enable higher achievable rates. The average values of $R_{\max}$ and $R_{\mathrm{th}}$ in bps/Hz for the considered four cases $\left(\theta,\sigma^2\right)=\{\left(0.05,\right.$ $\left.-70\text{ dBm}\right),\left(0.1,-70 \text{ dBm}\right),\left(0.05,-100 \text{ dBm}\right),\left(0.1,-100\right.$ $\left.\text{dBm}\right)\}$ are given by $\{106.60,114.42,146.47,154.28\}$ and $\{17.46,19.05,26.92,28.73\}$, respectively. When $R<R_{\mathrm{th}}$, the maximum received RF power $P_{R,E}^*$ for EH is achieved with EB. However, as $R$ increases and becomes substantially larger than $R_{\mathrm{th}}$, $P_{R,E}^*$ decreases till reaching a minimum value. For the latter cases, SM is adopted to achieve rate $R$ and the remaining received power is used for EH. 
 
The corresponding variation of the optimal power allocation $p_1^*$, $p_2^*$, $p_3^*$, and $p_4^*$ over the $r=4$ available eigenchannels is depicted in Fig$.$~\ref{fig:combined}(b) as a function of $R$ and $\theta$ for $\sigma^2=-70$dBm. To meet the increasing rate requirement $R$, the optimal power $p_1^*$ over the best gain eigenchannel monotonically decreases from $p_1^*{\,\cong\,} P_T$ (this happens for $R\le R_{\mathrm{th}}$ when EB is adopted) to equal power allocation $p_1^*{\,\cong\,} p_2^*{\,\cong\,}\frac{P_T}{r}$ for $R\approxeq R_{\max}$ (when SM needs to used). 
  
In Fig.~\ref{fig:combined}(c), the optimal UPS ratio $\rho^*$ is plotted versus $R$ for $2\times 2$ and $4\times 4$ MIMO systems. It is shown that $\rho^*$ monotonically decreases with increasing $R$ in order to ensure that sufficient fraction of the received RF power is used for ID, thus, to satisfy the rate requirement. Lower $\sigma^2$, larger $N$, and higher $\theta$ result in meeting $R$ with lower fraction $1-\rho$ of the received RF power dedicated for ID. Thus, for these cases, larger portion of the received RF power can be used for EH.

\begin{figure}[!t]
\centering
\subfigure[$2\times2$  MIMO system.]{{\includegraphics[width=1.7in]{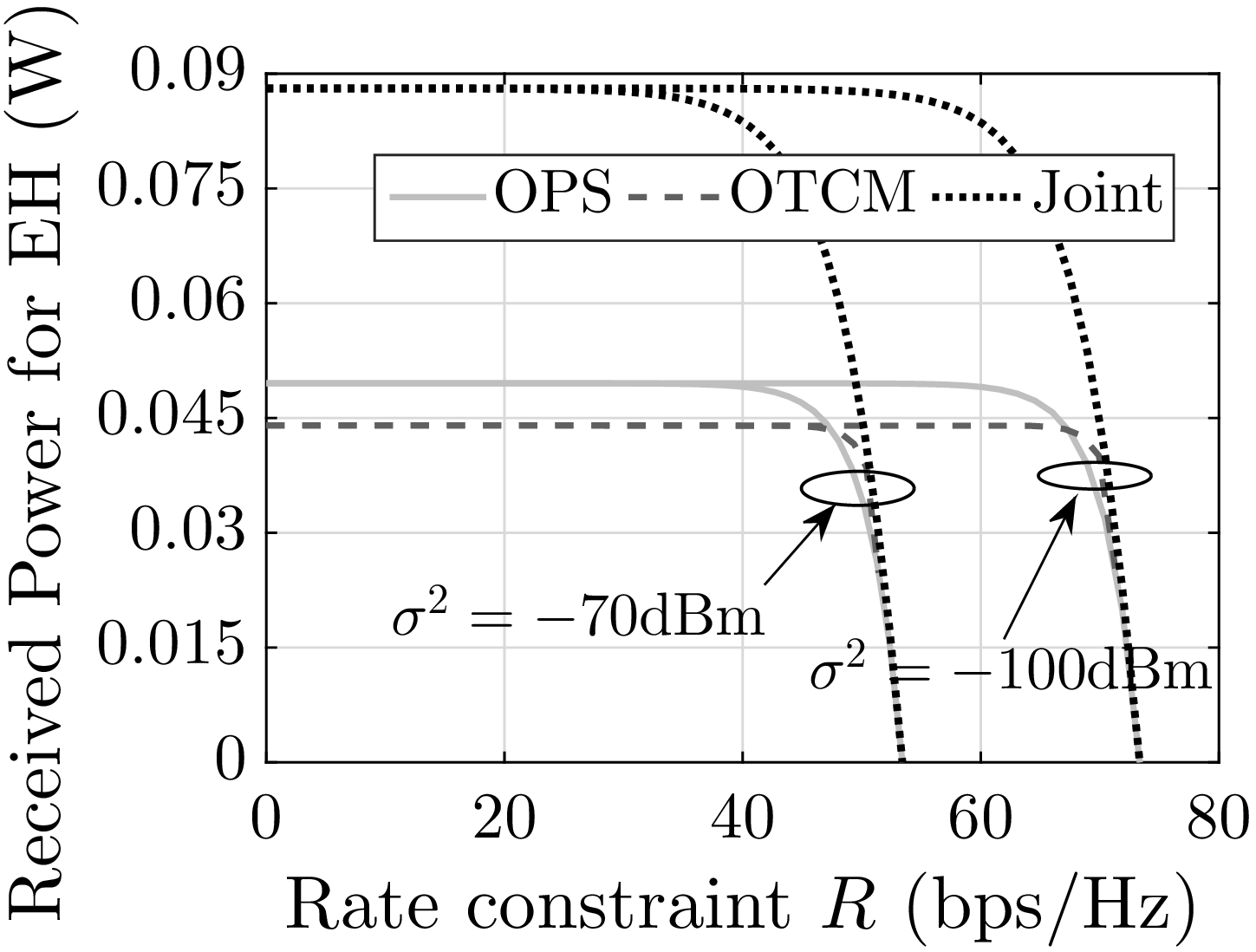} }}\hspace{-1.5mm}
\subfigure[$4\times4$  MIMO system.]
{{\includegraphics[width=1.7in]{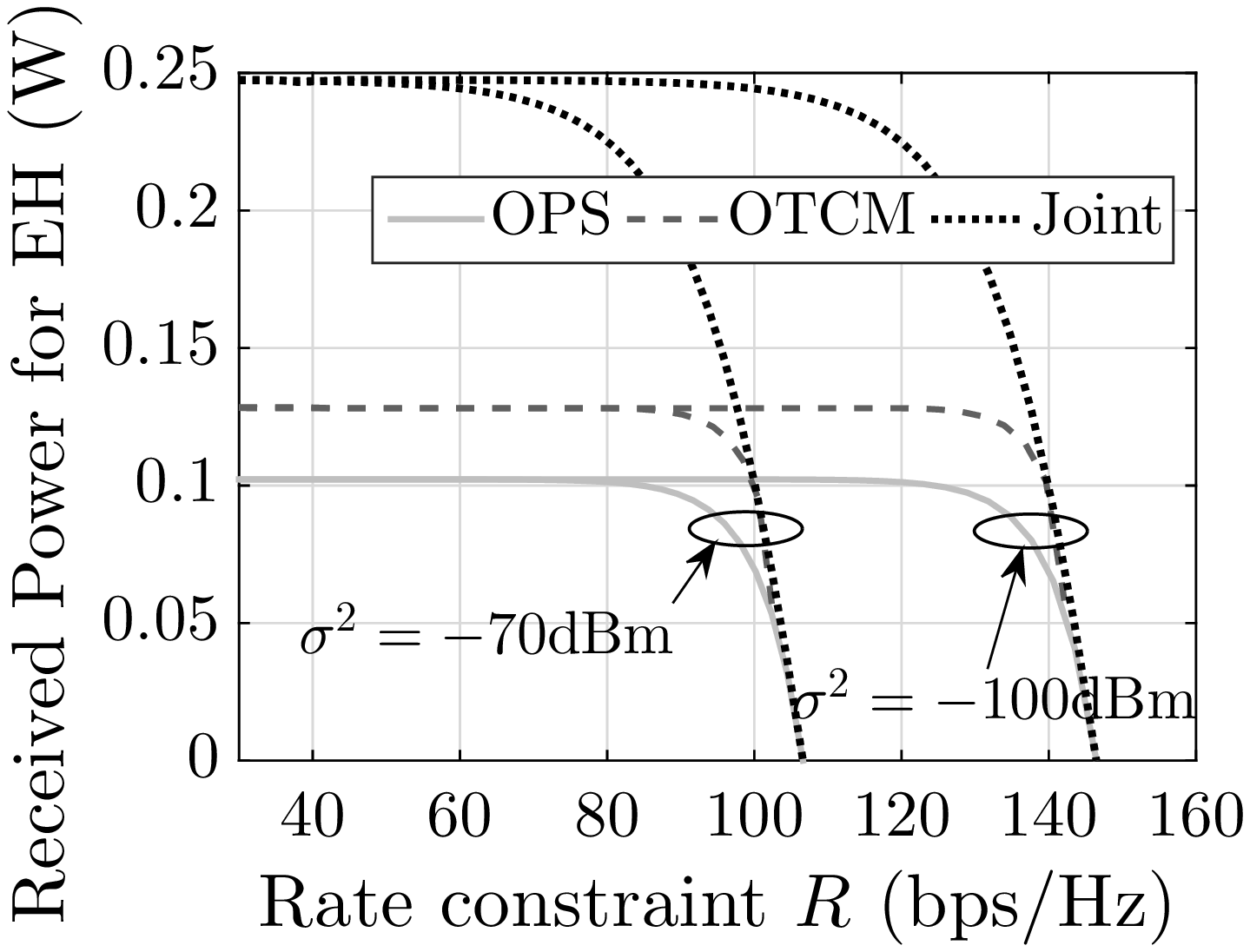} }}
\caption{\small Comparison of the rate-energy trade off between the proposed design and the benchmark semi-adaptive schemes OPS and OTCM.}\label{fig:comp}
\end{figure}

We finally present in Fig.~\ref{fig:comp} performance comparison results between the proposed joint TX precoding and RX UPS design and two benchmark schemes to highlight the importance of our considered joint optimization framework. The first scheme, termed as Optimal TX Covariance Matrix (OTCM), performs optimization of the TX covariance matrix $\mathbf{S}$ for a fixed UPS ratio $\rho$, and the second scheme, termed as Optimal UPS Ratio (OPS), optimizes $\rho$ for given $\mathbf{S}$. It is observed that for $2\times2$ MIMO systems, OPS performs better than OTCM, while for $4\times4$ MIMO systems, the converse is true. This happens because OTCM performance improves with increasing $N$. For both $N$ values, the proposed joint TX and RX design provides significant energy gains over OTCM and OPS. This performance enhancement for $N=2$ is $71.15$\% and $87.4$\%, respectively, over OPS and OTCM schemes, while it becomes $127$\% and $77.4$\%, respectively, for $N=4$.

\section{Conclusion}\label{sec:conclusion} 
In this paper we first prove the generalized convexity property of the harvested power maximization problem in QoS-constrained MIMO SWIPT systems with UPS reception and generic practical RF EH modeling. We next derive the global jointly optimal TX precoding and RX UPS ratio design. Different from recently proposed designs, the presented design solutions unveiled that there exists a rate requirement value that determines whether the TX precoding operation is EB or information SM. Representative numerical investigations showcased that our joint design results in nearly doubling the harvested power compared to benchmark schemes, thus enabling efficient MIMO SWIPT communication. We intend to extend our optimization framework in multiuser MIMO systems and consider nonuniform PS reception in future works.

\section*{Acknowledgments}
This work was supported by the 2016 Raman Charpak fellowship program. The views of Dr. Alexandropoulos expressed here are his own and do not represent Huawei's ones.
 
\bibliographystyle{IEEEtran}
\bibliography{refs_MIMO_PS_TBF}
\end{document}